\hfuzz 2pt
\font\titlefont=cmbx10 scaled\magstep1
\magnification=\magstep1

\null
\vskip 1.5cm
\centerline{\titlefont COMPLETE POSITIVITY}
\medskip
\centerline{\titlefont AND NEUTRON INTERFEROMETRY}
\vskip 2.5cm
\centerline{\bf F. Benatti}
\smallskip
\centerline{Dipartimento di Fisica Teorica, Universit\`a di Trieste}
\centerline{Strada Costiera 11, 34014 Trieste, Italy}
\centerline{and}
\centerline{Istituto Nazionale di Fisica Nucleare, Sezione di 
Trieste}
\vskip 1cm
\centerline{\bf R. Floreanini}
\smallskip
\centerline{Istituto Nazionale di Fisica Nucleare, Sezione di 
Trieste}
\centerline{Dipartimento di Fisica Teorica, Universit\`a di Trieste}
\centerline{Strada Costiera 11, 34014 Trieste, Italy}
\vskip 2cm
\centerline{\bf Abstract}
\smallskip
\midinsert
\narrower\narrower\noindent
We analyze the dynamics of neutron beams in interferometry 
experiments using quantum dynamical semigroups. 
We show that these experiments could provide stringent
limits on the non-standard, dissipative terms appearing
in the extended evolution equations.
\endinsert
\bigskip
\vfil\eject

An open quantum system can be modeled in general as
being a small subsystem in interaction with a suitable
large environment. Although the global dynamics of the 
compound system is described by unitary transformations
generated by the total hamiltonian, the effective time 
evolution of the subsystem usually manifests dissipation
and irreversibility.

This reduced dynamics, obtained by eliminating ({\it i.e.}
by tracing over) the environment degrees of freedom, turns 
out to be in general rather complicated. However, under
some mild assumptions, that essentially ask for a weak
coupling between system and environment, one obtains a
subdynamics free from memory effects, that can be realized in 
terms of linear maps. Furthermore, this set of transformations 
possesses very basic and fundamental physical properties, like
forward in time composition (semigroup property), probability
conservation, entropy increase and complete positivity.
They form a so-called quantum dynamical semigroup.[1-3]

This description is rather universal and can be applied
to model a large variety of different physical situations,
ranging from the study of quantum statistical systems,[1-3]
to the analysis of dissipative effects in quantum optics,[4-6]
to the discussion of the interaction of a microsystem with a 
macroscopic measuring apparatus.[7-9] 
It has also been proposed as an effective description for
phenomena leading to loss of quantum coherence,[10-14] induced by
quantum gravity effects at Planck's scale.[15]
The basic idea is that space-time should be topologically
nontrivial at this scale, manifesting a complicated ``foamy''
structure; as a consequence, transitions from pure to mixed
states could be conceivable (for more recent elaborations,
see [16]).

Analysis based on the study of the dynamics of strings also
support, from a different point of view, the idea of loss
of quantum coherence.[17,18] In this respect, one can show that,
in a rather model-independent way, a non-unitary, dissipative,
completely positive subdynamics is the direct result of the
weak interaction with an environment constituted by a gas of
D0-branes, effectively described by a heat-bath of quanta obeying
infinite statistics.[18]

These results suggest that it should be possible to observe 
dissipative, non-standard effects in many quantum systems.
However, crude dimensional estimates show that these effects 
are in general tiny, and therefore unobservable in practice.

Nevertheless, a detailed analysis of the dynamics and decay of
the neutral kaon system based on quantum dynamical semigroups
shows that the non-standard contributions in the extended
time-evolution equations could lead to testable effects.[19, 20, 21-23]
These non-standard terms can be parametrized by six real 
phenomenological constants, whose presence modify the expressions
of usual $K$-$\overline{K}$ observables, like decay rates 
and asymmetries.[21, 22] Although the present experimental data are
not accurate enough to detect these modifications, the next generation
of kaon experiments should be able to put stringent limits
on the six non-standard parameters. In this respect,
particularly promising are the experiments at $\phi$-factories,[23]
where systems of correlated neutral kaons are copiously produced.
One should note that the description of the
kaon dynamics in terms of completely positive maps is in this
case essential in order to obtain a consistent extension of 
the standard quantum mechanical time-evolution (for a complete discussion,
see [24]).
\vfill\eject

Another system in which non-standard quantum evolutions based
on dynamical semigroups can be studied is a neutron interferometer.[25-28]
The capability of producing very slow neutron beams at reactors,
together with the technological ability of producing and cutting
with high precision macroscopic silicon crystals have made
possible direct, very accurate tests of various physical phenomena.[25-30]

In a typical experimental setup, a neutron beam is split into two
components which travel along different paths and are subsequently
brought together to interfere. The two components pass through a
tiny slab of material before interfering; this produce a relative
``phase shift'' between the two splitted beams. By varying the relative
orientation of the slab across the two beams, 
one obtains an interference figure.
This figure changes under the action of various external phenomena,
produced {\it e.g.} by earth gravity and rotation, 
or by an external magnetic field
(earth magnetic field is usually properly screened); the corresponding
theoretically calculated phase shifts induced by these phenomena have all
been experimentally checked with high precision by analyzing the
modified interference patterns.[28-30]

In the following, we shall analyze in detail the dynamics of the
neutron beams in such interferometric devices under the hypothesis
that the corresponding time-evolution be described by a quantum
dynamical semigroup. We shall see that, at least in principle,
neutron interferometry experiments could provide very accurate
estimates of the non-standard, dissipative terms appearing in
the corresponding evolution equations. A preliminary analysis based
on recent published data from one of those experiments will
also be presented.

\vskip 1cm

States of a quantum system evolving in time can be suitably described
by a density matrix $\rho$; this is a positive, hermitian operator,
{\it i.e.} with positive eigenvalues, and constant trace.
We shall analyze the evolution of neutrons in an abstract interferometer, 
where the original monoenergetic beam is splitted in two components 
that then interfere at the end, giving rise to intensity fringe
patterns in two possible exit beams. We can model this generic
physical setup by means of a two-dimensional Hilbert space, taking
as basis states those corresponding to the two components 
of the split beam inside the interferometer.

With respect to this basis, the density matrix $\rho$ describing the
state of our physical system can be written as:
$$
\rho=\left(\matrix{
\rho_1&\rho_3\cr
\rho_4&\rho_2}\right)\ , \eqno(1)
$$
where $\rho_4\equiv\rho_3^*$, and $*$ signifies complex conjugation.

As explained in the introductory remarks, our analysis is based
on the assumption that the evolution in time of the neutrons inside
the interferometer is given by a quantum dynamical semigroup,
{\it i.e.} by a completely positive, one parameter
(=time) family of linear maps: $\rho(0)\mapsto\rho(t)$.
These maps are generated by equations of the following form:
$$
{\partial\rho(t)\over\partial t}=-iH \rho(t)+i\rho(t) H 
+L[\rho(t)] .\eqno(2)
$$
The first two terms in the r.h.s. of this equation are the standard
quantum mechanical ones. They contain the effective (time-independent)
hamiltonian $H$, that can be taken to be hermitian, since the fact
that the neutrons are unstable can be neglected in interferometry
experiments.

The third piece $L[\rho]$ is a linear map, whose form is fully 
determined by the requirement of complete positivity and trace
conservation:[1-3]
$$
L[\rho]=-{1\over2}\sum_j\Big(A^\dagger_jA_j\,\rho +
\rho\, A^\dagger_jA_j\Big)\ 
+\sum_j A_j\,\rho\, A^\dagger_j\ . \eqno(3)
$$
The operators $A_j$ must be such that $\sum_j A^\dagger_j A_j$ is a
well-defined $2\times 2$ matrix; further, to assure
entropy increase, the $A_j$ can be taken to be hermitian.
In absence of $L[\rho]$, pure states ({\it i.e.} states of the
form $|\psi\rangle\langle\psi|$) would be transformed into
pure states. Instead, the additional piece in (3) produces
dissipation and possible transitions from pure to mixed
states. 

As already mentioned, equations of the form (2), (3) have been 
used to describe various phenomena connected with open 
quantum systems; in particular, they have been 
applied to analyze the propagation and decay of the
neutral kaon system.[21-24] Although the basic general idea
behind these treatments is that quantum phenomena at
Planck's length produce loss of phase-coherence,
it should be stressed that the form (2), (3) of the evolution
equation is independent from the microscopic mechanism
responsible for the dissipative effects. Indeed, equations
(2) and (3) are the result of very basic physical assumptions, 
like probability conservation, entropy increase, complete
positivity, and therefore should be regarded as 
phenomenological in nature.

Among the just mentioned physical requirements, complete positivity
is perhaps the less intuitive. Indeed, it has not been
enforced in previous analysis, in favor of the more obvious
simple positivity.[10, 19, 20] 
Simple positivity is in fact enough to guarantee that
the eigenvalues of the density matrix $\rho(t)$ describing our
system remain positive at any time; this is an unavoidable requirement
in view of the probabilistic interpretation of $\rho$.

Complete positivity is a stronger property, in the sense that
it assures the positivity of the density matrix describing the
states of a larger system, obtained by coupling in a trivial
way the system under study with another arbitrary 
finite-dimensional one. At first, the requirement of complete
positivity of (2) seems a mere technical complication.
Nevertheless, it turns out to be essential in properly
treating correlated systems, like the two neutral kaons coming
from the decay of a $\phi$-meson; it assures the absence of 
unphysical effects, like the appearance of negative probabilities,
that could occur for just simply positive dynamics.[23, 24]
One should also add that standard unitary quantum mechanical time evolutions
satisfies this property in a rather trivial way.
For these reasons, in analyzing possible non-standard, dissipative
effects even in simpler, non correlated systems, 
the phenomenological equations (2) and (3) should be used.

In the particular case of neutron interferometers, as for
the $K$-$\overline{K}$ system, a more explicit description of
(2) and (3) can be given. In the chosen basis, the effective
hamiltonian can be written as:
$$
H=\left(\matrix{
E+\omega&0\cr
0&E-\omega\cr}\right)\ . \eqno(4)
$$
Indeed, the neutron beams inside the interferometer can be
assimilated to a two-level system, with $E$ the incident 
(kinetic) neutron energy. The splitting in energy $2\,\omega$
among the two internal beams can be induced by various 
physical effects. In the following, we shall consider the case 
of a thin slab of material ({\it e.g.} alluminium) inserted
transversally to the two split beams. A slight rotation of this
slab produces different effective interactions of the neutrons
with the slab material in the two internal paths, yielding
a non-vanishing $\omega$. Using an eikonal approximation,
perfectly suitable for describing slow neutrons, one can
theoretically compute this energy splitting in terms of 
the neutron-nuclear scattering parameters for the slab material.[25-27]
Typically, one finds that $\omega$ is of the order of
\hbox{$10^{-7}$ eV}.

The explicit form of the term $L[\rho]$ in (3) can be most 
simply given by expanding the $2\times2$ matrix $\rho$
in terms of Pauli matrices $\sigma_i$ and the identity $\sigma_0$:
$\rho=\rho_\mu\, \sigma_\mu$, $\mu=\,0$, 1, 2, 3.
In this way, the map $L[\rho]$ can be represented by
a symmetric $4\times 4$ matrix $\big[L_{\mu\nu}\big]$, 
acting on the column vector with components $(\rho_0,\rho_1,\rho_2,\rho_3)$.
It can be parametrized by the six real 
constants $a$, $b$, $c$, $\alpha$, $\beta$, and $\gamma$:[21]
$$
\big[L_{\mu\nu}\big]=-2\left(\matrix{0&0&0&0\cr
                                     0&a&b&c\cr
                                     0&b&\alpha&\beta\cr
                                     0&c&\beta&\gamma\cr}\right)
\ ,\eqno(5)
$$
with $a$, $\alpha$ and $\gamma$ non-negative.
These parameters are not all independent; the condition of
complete positivity of the time-evolution 
$\rho\rightarrow\rho(t)$ imposes the following inequalities:
$$
\eqalign{
&2R\equiv\alpha+\gamma-a\geq0\ ,\cr
&2S\equiv a+\gamma-\alpha\geq0\ ,\cr
&2T\equiv a+\alpha-\gamma\geq0\ ,\cr
&RST\geq 2\, bc\beta+R\beta^2+S c^2+T b^2\ .
}\hskip -1cm
\eqalign{
&RS\geq b^2\ ,\cr
&RT\geq c^2\ ,\cr
&ST\geq\beta^2\ ,\cr
&\phantom{\beta^2}\cr
}\eqno(6)
$$

As already observed, the dissipative correction (5) to the
evolution equation (2) should be regarded as phenomenological;
it is therefore difficult to give an apriori estimate of the
magnitude of the non-standard parameters in (5). However,
following the idea that the term $L[\rho]$ originates from
quantum effects at Planck's scale, one expects the values
of $a$, $b$, $c$, $\alpha$, $\beta$ and $\gamma$ to be very small,
at most of the order $m_n^2/m_P\simeq 10^{-19}\ {\rm GeV}$,
where $m_n$ is the neutron mass, while $m_P$ is the Planck scale.
The dissipative contribution $L[\rho]$ in (2) is therefore at least
three order of magnitude smaller than the one given by the standard
hamiltonian terms: this allows an approximate analysis of the
evolution equation (2), in which the term (5) can be treated as a small
perturbation.

For the considerations that follows, it will be sufficient to stop
at the first order in the perturbative expansion. For generic initial
conditions, the time dependence of the four components
of the corresponding solution $\rho(t)$ of (2) is explicitly given by:
$$
\eqalignno{
&\rho_1(t)=(1-\gamma t)\, \rho_1 + \gamma t\, \rho_2 
-{C\over\omega}\, e^{-i\omega t}\,\sin(\omega t)\ \rho_3
-{C^*\over\omega}\, e^{i\omega t}\, \sin(\omega t)\ \rho_4\ ,&(7a)\cr
&\rho_2(t)=\gamma t\, \rho_1 + (1-\gamma t)\, \rho_2 
+{C\over\omega}\, e^{-i\omega t}\, \sin(\omega t)\ \rho_3
+{C^*\over\omega}\, e^{i\omega t}\, \sin(\omega t)\ \rho_4\ ,&(7b)\cr
&\rho_3(t)=-{C^*\over\omega}\, e^{-i\omega t}\,\sin(\omega t)\ (\rho_1-\rho_2)
+(1-At)\, e^{-2i\omega t}\, \rho_3 + {B\over 2\omega}\,\sin(2\omega t)\ \rho_4
\ ,&(7c)\cr
&\rho_4(t)=-{C\over\omega}\, e^{i\omega t}\,\sin(\omega t)\ (\rho_1-\rho_2)
+{B^*\over 2\omega}\,\sin(2\omega t)\ \rho_3+(1-At)\, e^{2i\omega t}\, \rho_4 
\ ,&(7d)\cr
}
$$
where the following convenient combinations of the non-standard
parameters have been introduced:
$$
A=\alpha+a\ ,\quad B=\alpha-a+2ib\ ,\quad C=c+i\beta\ .\eqno(8)
$$

Any physical property of the diffracted neutron beams exiting the
interferometer can be extracted from the solution (7) for the
density matrix $\rho(t)$ by computing its trace with suitable
hermitian operators. In particular, the observation of the
neutron intensity pattern just outside the interferometer
corresponds to the computation of the mean value of the following
projector operators:[10]
$$
{\cal O}_+={1\over2}\left(\matrix{
1&e^{i\theta}\cr
e^{-i\theta}&1\cr
}\right)\ ,\qquad\quad
{\cal O}_-={1\over2}\left(\matrix{
1&e^{i(\theta+\pi)}\cr
e^{-i(\theta+\pi)}&1\cr
}\right)\ ,\eqno(9)
$$
that refer to the two possible exit beams in which a neutron 
can be found, having traveled the whole interferometer;[27]
the parameter $\theta$ is a phase which depends on the specific
experimental setup. Then, the intensity $I_\pm$ of the interference
figure in the two exit beams is given by: 
$$
I_\pm(t)=\langle{\cal O}_\pm\rangle={\rm Tr}[{\cal O}_\pm\,\rho(t)]\ .
\eqno(10)
$$

In the basis we are using, the initial conditions for a neutron
entering the interferometer is given by one of the following
two density matrices:
$$
\rho^{(1)}={1\over2}\left(\matrix{
1&1\cr
1&1\cr
}\right)\ ,\qquad
\rho^{(2)}={1\over2}\left(\matrix{
\phantom{-}1&-1\cr
-1&\phantom{-}1\cr
}\right)\ .\eqno(11)
$$
They correspond to the two possible choices of orientation of the incident
neutron beam with respect to the interferometer, and give rise to the
same final results; in the following, we shall work with $\rho^{(1)}$.
Inserting this initial condition in the time evolution given by (7),
from (10) one obtains the following two interference patterns:
$$
I_\pm(t)={1\over2}\bigg\{1 \pm\bigg[e^{-At}\cos\big(\theta+2\omega t\big) +
{|B|\over 2\omega}\, \sin(2\omega t)\, \cos(\theta-\theta_B)\bigg]\bigg\}\ ,
\eqno(12)
$$
where $|B|$ and $\theta_B$ are modulus and phase of $B$ in (8);
this formula holds for times such that $At\ll 1$.
Since a neutron having traveled fully inside the interferometer
can only be detected in one of the two exit beams,
particle conservation requires: $I_+(t) +I_-(t)=1$, as it is evident
from (12).

The interference figures described by (12) are those produced
by a perfectly monoenergetic neutron incident beam and an
ideal interferometer. In practice, the neutron momenta in the
incident beam have a finite distribution of magnitude and directions;
furthermore, there are always slight imperfections in the
construction of the interferometer, not to mention
residual strains in the crystal itself. These effects can only partially
be controlled, and produce significant attenuation in the intensity of the
interference figures.
Detailed calculations based on neutron optics allow
precise estimates of the modifications that are needed in the
spectra (12) in order to take into account those effects.[25-27]
In keeping with our phenomenological point of view, we will not
use directly those estimates, but rather modify the expressions (12) 
by introducing suitable unknown parameters.
By denoting with $N_\pm$ the actual neutron countings at the two
exit beams, one generalizes the spectra in (12) as:
$$
N_\pm(t)=N^{(0)}_\pm\,\bigg\{1 \pm {\cal C}_\pm\,
\bigg[e^{-At}\cos\big(\theta+2\omega t\big) +
{|B|\over 2\omega}\, \sin(2\omega t)\, \cos(\theta-\theta_B)\bigg]\bigg\}\ .
\eqno(13)
$$
The parameters ${\cal C}_\pm$, the so-called fringe contrast, take
into account the previously mentioned intensity attenuation,
while $N^{(0)}_\pm$ are just normalization constants.
Clearly, the accuracy of the determination of the parameters 
$A$ and $B$ from the measured data will increase 
as the fringe contrast gets closer to one.
In actual experiments, one finds that the best values
for ${\cal C}_\pm$ are usually around 0.6.
Further, note that now particle conservation requires:
$$
N^{(0)}_+\, {\cal C}_+=N^{(0)}_-\, {\cal C}_-\ .\eqno(14)
$$

In order to be able to compare the phenomenological predictions (13)
with actual experimental data, further elaborations are required.
Although the intensity spectra in (13) are time-dependent, 
in an interference experiment, being the paths followed by the neutrons
fixed, the evolution time $t$ is also fixed; it can only be
modified by changing the wavelength ({\it i.e.} the energy) 
of the primary neutron beam. It follows that the phase
$\varphi\equiv 2\,\omega\, t$ that gives the interference figure
can be varied only by changing the energy split $\omega$ between the two
paths inside the interferometer, {\it i.e.} by changing the
orientation of the material slab with respect to the neutron beams.

Also, since $t$ is fixed, it is not possible a priori to extract
the fringe contrast parameters ${\cal C}_\pm$ from the $\varphi$-dependence
of the intensities $N_\pm$. In other words, in comparing the behaviour
in (13) with that given by the experiment, one needs to use the following
form for the two intensity patterns:
$$
N_\pm(\varphi)=N^{(0)}_\pm\,\bigg\{1 \pm 
\bigg[
P_\pm\,\cos\big(\theta+\varphi\big) +
Q_\pm\, {\sin\varphi\over\varphi}\bigg]\bigg\}\ ,
\eqno(15)
$$
where
$$
P_\pm={\cal C}_\pm\, e^{-At}\ ,\qquad 
Q_\pm={\cal C}_\pm\, |B| t\, \cos\big(\theta-\theta_B\big)\ .
\eqno(16)
$$
A fit with the experimental data will give estimates for
the parameters $N^{(0)}_\pm$, $P_\pm$, $Q_\pm$ and $\theta$;
at least in principle, this is sufficient to determine 
the non-standard constants $A$ and $B$.

We have performed a preliminary $\chi^2$ fit of the formulas
in (15) with recent experimental data, published in [30].
In that experiment, a so-called skew-symmetric silicon
interferometer and polarized neutrons were used to study
the ``geometrical phase'' of the neutron wavefunction;
however, in order to check the apparatus, also standard
interferometric spectra had been taken. In this way,
the experimental setup described in [30] is an actual realization
of the abstract interferometer so far discussed in deriving (15).
Although the results of other recent interferometry 
experiments are also available,[28] in lacking for an ad-hoc
experiment, we find the presentation of the data in [30] the
more suitable for the analysis of the consequences of the
evolution equation (2) and (5).
The results of our fit can be summarized as follows:
$$
\eqalignno{&N^{(0)}_+=942\pm6\qquad
P_+=0.17\pm0.01\qquad
Q_+=0.02\pm0.02\qquad
\theta=0.09\pm0.05\qquad &(17a)\cr
&N^{(0)}_-=366\pm4\qquad
P_-=0.46\pm0.02\qquad
Q_-=0.06\pm0.02\qquad
\theta=0.03\pm0.03\ .&(17b)}
$$

Extracting estimates for the parameters $A$ and $B$ from
these results requires the determination of the fringe
contrast ${\cal C}_\pm$, characterizing the ``optical'' properties
of the neutron interferometer. These constants are in fact
function of the imaginary part of the refraction index
of the material from which the interferometer is built.[27]
Therefore, the best way to obtain
estimates for ${\cal C}_\pm$ is to measure the interference spectra
with two neutron beams of wavelength $\lambda$ and $\lambda/2$,[29]
keeping unchanged the rest of the experimental setup.
By comparing the corresponding fit estimates for the
coefficients $P_\pm$ and $Q_\pm$, with the help of 
neutron optics theory, one is then able
to extract the values of ${\cal C}_\pm$ at wavelength $\lambda$.

In lacking of a two-wavelength experiment, in the following
we shall estimate ${\cal C}_\pm$ using directly the data.
In the standard quantum mechanical case, {\it i.e.} for
$A=B=\,0$, one can easily obtain the coefficients ${\cal C}_\pm$ 
from the maximum $N^{(\rm max)}$ and the minimum
$N^{(\rm min)}$ neutron counts 
of the experimental interference figures.
Indeed, from (15) with $A=B=\,0$, one obtains:
$$
{\cal C}_\pm={N_\pm^{(\rm max)}-N_\pm^{(\rm min)}\over
N_\pm^{(\rm max)}+N_\pm^{(\rm min)}}\ .\eqno(18)
$$
Although this relation is only approximately valid for nonvanishing
$A$ and $B$, in practice one can still use (18) with confidence, 
since the systematic error that one thus makes in the evaluation of the
parameters $A$ and $B$ can be estimated at the end to be 
much smaller than the pure experimental uncertainty.

Using the experimental data and (18), one obtains:
${\cal C}_+=0.19\pm0.02$ and ${\cal C}_-=0.54\pm0.03$. As an independent
test of the correctness of this evaluation, one can check
that, within the errors, the relation (14) is perfectly satisfied.
Note that the value of ${\cal C}_+$ is significally smaller than one,
while that of ${\cal C}_-$ is close to the best figures that can be
attained in practice.[26-28] This difference in the fringe contrast
of the two data samples will result in a significally less accurate
determination of the non-standard parameters $A$ and $B$ from the
the results in $(17a)$ with respects to those in $(17b)$.

The flight time $t$ of the neutrons inside the interferometer
can be very accurately determined by using the geometric
specifications of the silicon interferometer used in [30],
so that $1/t=5.83\times10^{-21}\ {\rm GeV}$.
Further, on general grounds one expects ${\cal R}e(B)$ and 
${\cal I}m(B)$ to be of the same order of magnitude
(for the case of the neutral kaon system, see [22]); then, as a
working assumption, we shall 
neglect the small phase $\theta$ with respect to $\theta_B$
in (16), so that only the real part of $B$ can be extracted
from the estimates of $Q_\pm$. Putting everything together,
one finally obtains from $(17a)$:
$A=(0.71\pm0.73)\times 10^{-21}\ {\rm GeV}$
and ${\cal R}e(B)=(0.76\pm0.49)\times 10^{-21}\ {\rm GeV}$,
while from $(17b)$:
$A=(0.84\pm0.41)\times 10^{-21}\ {\rm GeV}$
and ${\cal R}e(B)=(0.65\pm0.24)\times 10^{-21}\ {\rm GeV}$.
These two estimates are compatible, but, as expected, the second one
is much more accurate.

Alternatively, recalling the definitions (8), one can express
the previous results as an estimate for the
parameters $a$ and $\alpha$ of (5); using the best values
for $A$ and ${\cal R}e(B)$, one finds: $a=(0.10\pm0.24)\times 10^{-21}\
{\rm GeV}$, $\alpha=(0.74\pm0.24)\times 10^{-21}\ {\rm GeV}$.
Although these values should be taken as indicative, they seem
to suggest a possible nonvanishing value for $\alpha$,
while $a$ is compatible with zero at the present level of
accuracy.

If the non-standard parameter $a$ actually vanishes,
the expression (5) of the extra term $L[\rho]$ in
the evolution equation (2) greatly simplifies.
Indeed, for $a=\,0$, the inequalities (6) readily imply:
$\gamma=\alpha$, $b=c=\beta=\,0$. In this case,
the evolution equation (2) gives the most simple extension
of ordinary quantum mechanics, compatible with the
condition of complete positivity.

In this simplified situation, the combinations $A$ and $B$ in (8)
become both equal to $\alpha$, so that the relations (16) are modified as:
$$
P_\pm={\cal C}_\pm\, e^{-\alpha t}\simeq {\cal C}_\pm\, (1-\alpha t)\ ,\qquad
Q_\pm={\cal C}_\pm\, \alpha t\, \cos\theta\ .\eqno(19)
$$
By eliminating $\alpha t$ from these two formulas, one is now able
to determine the fringe contrast factors ${\cal C}_\pm$ from the fitted
parameters $P_\pm$ and $Q_\pm$ without further assumptions:
${\cal C}_+=0.20\pm0.02$, ${\cal C}_-=0.52\pm0.03$. Note that these values are
equal within errors to those determined before.
Then, from either of the relations in (19), one obtains two
determinations of the parameter $\alpha$, which
combined finally give:
$$
\alpha=(0.71\pm0.21)\times 10^{-21}\ {\rm GeV}\ .\eqno(20)
$$
Although this estimate for $\alpha$ points toward a nonvanishing value,
roughly of the right order of magnitude for a quantum gravity
or ``stringy'' origin, it should not be regarded as an evidence
for non-standard, dissipative effects in the dynamics describing
neutron interferometry. Rather, it should be considered as a rough
evaluation of the sensitivity that present neutron interferometry
experiments can reach in testing quantum dynamical time
evolutions of the form given in (2) and (5).

In closing, we would like to make a few comments on the
existing literature on the subject.
In our study of the effects of the environment on the
propagation of the neutrons inside the interferometer, 
we have assumed that the refractive phenomena on the various
silicon blades of the device be described by standard neutron
optics. This theory is the result
of a quantum mechanical analysis of the scattering
of the neutron beams by the nuclei in the silicon crystal.
For slow neutrons, the effects of these scattering phenomena can
be effectively described by a model that resemble very closely
``geometric optics'' of light propagation theory.[25-27]

In principle, non-standard dissipative effects can also be 
present in the scattering neutron-nucleus;[31-33] these phenomena
can be described again by phenomenological equations of the
form (2), (3), and could modify the predictions of standard
neutron optics. A precise estimate of these changements 
requires detailed computations that certainly go beyond the
scope of the present investigation. In any case, it should be 
stressed that these possible dissipative effects in the interaction
neutron-nucleus would mostly affect the estimate of the
intrinsic interferometer parameters, like the fringe contrast
or the phase $\theta$, as functions of the
wavelength of the incident neutrons and the nuclear
properties of the refractive material. In our study, these
parameters have been obtained directly from the experimental data,
so that we expect little changements in our analysis from these
extra effects. Nevertheless, the entire topic certainly deserves
further attention and we hope to come back to these problems in
the future.

A study of possible phenomena violating quantum mechanics in neutron
interferometry has been originally presented in [10]. There, an equation
of the form (2) has also been used to describe these effects, but
without imposing the condition of complete positivity. Fitting an
approximated formula for the exit beams interference figures
with the experimental data available at that time, 
limits on some of the parameters violating quantum mechanics were
given. These limits have been further strengthen by later analysis,[34] 
exploiting wavefront splitting interference experiments (the analogs
of Young's two slit experiment).

These estimates turn out to be rather crude: they are based on a rough
evaluation of the flight-time of the neutrons inside the interferometer,
rather than a detailed analysis of the interference patterns.
Furthermore, as already mentioned, lacking of imposing the condition of
complete positivity on the evolution equation could lead to serious
inconsistencies.[24] We stress that to avoid these problems, one needs
to adopt phenomenological descriptions based on equations (2) and (3).

Finally, the neutron interference experiments realized so far 
allows determining at best the values of only two of the six 
non-standard parameters in (5). The remaining ones could be estimated,
at least in principle, by studying the behaviour of other
observables $\cal O$, different from those appearing in (9). 
This would allow a direct test of the inequalities (6) 
and therefore of the hypothesis of complete positivity.

In practice, however, the analysis of these new observables would
correspond to the realization of completely different experimental
setups. In this respect, a detailed study of possible 
non-standard, dissipative effects in neutron interferometry appears
to be an exiting challenge not only theoretically, but
experimentally as well.

\vfill\eject

\centerline{\bf REFERENCES}
\bigskip\medskip

\item{1.} R. Alicki and K. Lendi, {\it Quantum Dynamical Semigroups and 
Applications}, Lect. Notes Phys. {\bf 286}, (Springer-Verlag, Berlin, 1987)
\smallskip
\item{2.} V. Gorini, A. Frigerio, M. Verri, A. Kossakowski and
E.C.G. Surdarshan, Rep. Math. Phys. {\bf 13} (1978) 149 
\smallskip
\item{3.} H. Spohn, Rev. Mod. Phys. {\bf 53} (1980) 569
\smallskip
\item{4.} W.H. Louisell, {\it Quantum Statistical Properties of Radiation},
(Wiley, New York, 1973)
\smallskip
\item{5.} C.W. Gardiner, {\it Quantum Noise} (Springer, Berlin, 1992)
\smallskip
\item{6.} M.O. Scully and M.S. Zubairy, 
{\it Quantum Optics} (Cambridge University Press, Cambridge, 1997)
\smallskip
\item{7.} L. Fonda, G.C. Ghirardi and A. Rimini, Rep. Prog. Phys.
{\bf 41} (1978) 587 
\smallskip
\item{8.} H. Nakazato, M. Namiki and S. Pascazio,
Int. J. Mod. Phys. {\bf B10} (1996) 247
\smallskip
\item{9.} F. Benatti and R. Floreanini, Phys. Lett. {\bf B428} (1998) 149
\smallskip
\item{10.} J. Ellis, J.S. Hagelin, D.V. Nanopoulos and M. Srednicki,
Nucl. Phys. {\bf B241} (1984) 381; 
\smallskip
\item{11.} S. Coleman, Nucl. Phys. {\bf B307} (1988) 867
\smallskip
\item{12.} S.B. Giddings and A. Strominger, Nucl. Phys. {\bf B307} (1988) 854
\smallskip
\item{13.} M. Srednicki, Nucl. Phys. {\bf B410} (1993) 143
\smallskip
\item{14.} L.J. Garay, Phys. Rev. Lett. {\bf 80} (1998) 2508;
Thermal properties of spacetime foam, {\tt gr-qc/9806047}
\smallskip
\item{15.} S. Hawking, Comm. Math. Phys. {\bf 87} (1983) 395; Phys. Rev. D
{\bf 37} (1988) 904; Phys. Rev. D {\bf 53} (1996) 3099
\smallskip
\item{16.} S. Hawking and C. Hunter, Gravitational entropy and global
structure,\hfill\break {\tt hep-th/9808085}
\smallskip
\item{17.} J. Ellis, N.E. Mavromatos and D.V. Nanopoulos, Phys. Lett.
{\bf B293} (1992) 37; Int. J. Mod. Phys. {\bf A11} (1996) 1489
\smallskip
\item{18.} F. Benatti and R. Floreanini, Non-standard neutral kaons dynamics
from D-branes statistics, {\tt hep-th/9811196}
\smallskip
\item{19.} J. Ellis, J.L. Lopez, N.E. Mavromatos 
and D.V. Nanopoulos, Phys. Rev. D {\bf 53} (1996) 3846
\smallskip
\item{20.} P. Huet and M.E. Peskin, Nucl. Phys. {\bf B434} (1995) 3
\smallskip
\item{21.} F. Benatti and R. Floreanini, Nucl. Phys. {\bf B488} (1997) 335
\smallskip
\item{22.} F. Benatti and R. Floreanini, Phys. Lett. {\bf B401} (1997) 337
\smallskip
\item{23.} F. Benatti and R. Floreanini, Nucl. Phys. {\bf B511} (1998) 550
\smallskip
\item{24.} F. Benatti and R. Floreanini, 
Mod. Phys. Lett. {\bf A12} (1997) 1465; 
Banach Center Publications, {\bf 43} (1998) 71; 
Comment on ``Searching for evolutions 
of pure states into mixed states in the two-state system $K$-$\overline{K}$'',
{\tt hep-ph/9806450}
\smallskip
\item{25.} J.L. Staudenmann, S.A. Werner, R. Colella and A.W. Overhauser,
Phys. Rev. A {\bf 21} (1980) 1419
\smallskip
\item{26.} S.A. Werner and A.G. Klein, Meth. Exp. Phys. {\bf A23} (1986) 259
\smallskip
\item{27.} V.F. Sears, {\it Neutron Optics}, (Oxford University Press, Oxford,
1989)
\smallskip
\item{28.} {\it Advance in Neutron Optics and Related Research Facilities},
M. Utsuro, S. Kawano, T. Kawai and A. Kawaguchi, eds.,
J. Phys. Soc. Jap. {\bf 65}, Suppl.A, 1996
\smallskip
\item{29.} K.C. Littrell, B.E. Allman and S.A. Werner,
Phys. Rev. A {\bf 56} (1997) 1767
\smallskip
\item{30.} B.E. Allman, H. Kaiser, S.A. Werner, A.G. Wagh, V.C. Rakhecha
and J. Summhammer, Phys. Rev A {\bf 56} (1997) 4420
\smallskip
\item{31.} E.B. Davies, Ann. Inst. H. Poncar\'e, {\bf A 29} (1978) 395;
{\it. ibid.} {\bf A 32} (1980) 361
\smallskip
\item{32.} R. Alicki, Ann. Inst. H. Poncar\'e, {\bf A 35} (1981) 97;
Z. Phys. A {\bf 307} (1982) 279
\smallskip
\item{33.} L. Lanz and B. Vacchini, Int. J. Theor. Phys. {\bf 36} (1997) 67;
Phys. Rev. A {\bf 56} (1997) 4826
\smallskip
\item{34.} A.G. Klein, Phys. Lett. {\bf B151} (1985) 275;
Physica B {\bf 151} (1988) 44

\bye